\documentclass[letterpaper, 10 pt, conference]{ieeeconf}  % Comment this line out if you need a4paper

\IEEEoverridecommandlockouts                              % This command is only needed if 
\usepackage{graphicx} % for pdf, bitmapped graphics files
\usepackage{amsmath} % assumes amsmath package installed
\usepackage{amssymb}  % assumes amsmath package installed
\usepackage{cite}

\title{\LARGE \bf
Real-time Closed Loop Neural Decoding on a Neuromorphic chip
}

\author{Shoeb Shaikh$^{1}$, Rosa So$^{2}$, Tafadzwa Sibindi$^{3}$, Camilo Libedinsky$^{3}$ and Arindam Basu$^{1}$% <-this % stops a space
\thanks{**This work was supported through grant RG87/16 by MOE, Singapore}% <-this % stops a space
\thanks{$^{1}$Authors are with Nanyang Technological University, Singapore
        {\tt\small arindam.basu@ntu.edu.sg}}%
\thanks{$^{2}$Author is with the Institute for Infocomm Research, Singapore
        {\tt\small rosa-so@i2r.a-star.edu.sg}}%
\thanks{$^{3}$Authors are with SiNAPSE, National University of Singapore, Singapore
        {\tt\small camilo@nus.edu.sg}}%
}

\begin{document}
%\bstctlcite{IEEEexample:BSTcontrol}
\maketitle
\thispagestyle{empty}
\pagestyle{empty}

%%%%%%%%%%%%%%%%%%%%%%%%%%%%%%%%%%%%%%%%%%%%%%%%%%%%%%%%%%%%%%%%%%%%%%%%%%%%%%%%
\begin{abstract}

This paper presents for the first time a real-time closed loop neuromorphic decoder chip-driven intra-cortical brain machine interface (iBMI) in a non-human primate (NHP) based experimental setup. Decoded results show trial success rates and mean times to target comparable to those obtained by hand-controlled joystick. Neural control trial success rates of $\approx$ 96\% of those obtained by hand-controlled joystick have been demonstrated. Also, neural control has shown mean target reach speeds of $\approx$ 85\% of those obtained by hand-controlled joystick . These results pave the way for fast and accurate, fully implantable neuromorphic neural decoders in iBMIs.

\end{abstract}

%%%%%%%%%%%%%%%%%%%%%%%%%%%%%%%%%%%%%%%%%%%%%%%%%%%%%%%%%%%%%%%%%%%%%%%%%%%%%%%%
\section{INTRODUCTION}

Intra-cortical Brain Machine Interfaces (iBMIs) have made it possible for patients suffering from paralysis to establish communication \nocite{Pandarinath2017}, locomotion \cite{Rajangam2016} and even perform complex tasks such as feeding oneself \cite{Collinger2013, Ajiboye2017} solely through neural control. It is indeed heartwarming to see these chronic patients take a step towards leading normal lives as seen in demo videos presented in \cite{Collinger2013} and \cite{Ajiboye2017}. 

iBMI systems can be typically broken down into the following blocks – a) Signal acquisition and pre-processing, b) Feature extraction, c) Decoding and d) Effector. The signal acquisition block consists of a microelectrode array sampling extracellular neuronal electrical activity at around 30 - 40 kHz. This recorded activity data is conventionally passed on through wires/cables to bulky signal processing systems implementing blocks b), c) and d). These wires/cables leave an opening in the skull making the area prone to infection. They also impede the mobility of the system as a whole. To tackle the mobility issue, solutions such as \cite{Gao2012, Miranda2010, Chestek2009, Santhanam2007, Yin2014, Zanos2011, Schwarz2014} with wireless transmitter electronics outside the skin have been reported. However, these still have the presence of infection-prone transcutaneous wires. Meanwhile, mobile fully implantable and hermetically sealed prototypes with decreased risk of infection have been reported in \cite{Rizk2009,Borton2013,Harrison2007}. However, with exponentially rising number of electrodes \cite{Stevenson2011,Jun2017} and the concomitant increase in data rates, it is untenable to scale the idea of transmitting raw data in a fully implantable manner without heating up neural tissue excessively\cite{Kim2007,Dethier2011} or hitting bandwidth constraints\cite{Verma_DATE12}.

In order to solve this problem, we present a system level design space exploration in section II and advocate for performing operations up to decoding in the implant itself. There have been suggestions made along similar lines in the works reported in \cite{Dethier2013,Boi2016,Jiang2017,Rapoport2012,Chen2016}. However, \cite{Jiang2017,Rapoport2012,Chen2016} have reported offline results on pre-recorded data, of which \cite{Rapoport2012} has only implemented the proof of concept on an FPGA thus not presenting true energy benefits and also not showing live experimental results. \cite{Boi2016} has reported a real-time demonstration but the experiment has been performed on an anesthetized rat instead of the more standard use of awake non-human primates (NHP) and benchmarking has not been provided against state of the art decoding techniques. Authors in \cite{Dethier2013} have done an extensive study on standard experimental protocol of two NHPs in real-time. They have also benchmarked their algorithm against state of the art decoding technique, but the proposed decoder has been implemented in software leaving room for hardware demonstration. 

We present for the first time a real-time closed loop demonstration of a fully brain controlled neuromorphic decoder chip in an experimental setup involving an NHP. Previously, offline results comparing this decoder chip with state of the art technique were reported in \cite{Shaikh2017} in a similar experimental setup. We have benchmarked real-time brain controlled neuromorphic decoder's performance against hand-controlled joystick following the paradigm reported in \cite{Gilja2012}.

\section{IBMI$'$s - A SYSTEM LEVEL REVIEW}

\begin{figure}[h]
\centering
\includegraphics[width = 8.5 cm,height = 3 cm]{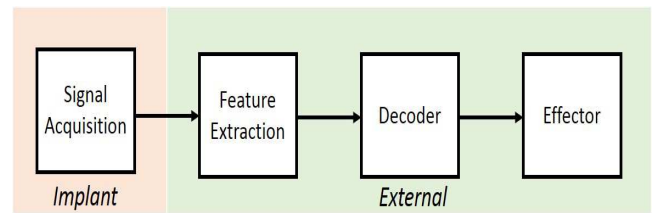}
\caption{Typical block diagram of an iBMI system}
      \label{fig:systemDiag}
\end{figure}

The state of the art iBMI systems can be illustrated in Fig. \ref{fig:systemDiag}.

If we consider a typical case of a $100$ electrode array with a sampling rate of $20$ kHz and raw data digitized at $12$ bits, the transmission rate becomes $24$ Mbps. A fully implantable solution reported for this transmission rate yields $\approx 7$ hours of continuous operation powered by a rechargeable Li-ion battery \cite{Borton2013}. Reported time for charging the battery is two hours, which makes daily use quite cumbersome and impractical. 
To eliminate this frequent recharging owed principally to the broadband data rate \cite{Harrison2008}, we propose incorporating the feature extraction and decoding blocks in the implant itself. 

\iffalse
\begin{figure}[h]
\centering
\includegraphics[width = 8.5 cm,height = 3 cm]{ProposedRecording.eps}
\caption{Block diagram of proposed iBMI system. }
      \label{fig:Systemdiag}
\end{figure}
\fi
Using this approach for a six degree of freedom robotic arm needing $10$ bit decoded outputs at $50$ Hz, data rate comes to $3$ kbps. This is approximately three orders of magnitude lower than transmitting the raw data.
Furthermore, the computational cost to include feature extraction block is $\approx 40$ nW per channel, which leads to a total of $4$ $\mu$W power consumption for $100$ electrodes. We propose ultra-low power neuromorphic \textbf{\underline{S}}pike-input \textbf{\underline{E}}xtreme \textbf{\underline{L}}earning \textbf{\underline{Ma}}chine (SELMA)  to be used as the decoder of choice since it yields $\approx 10$ times lower power consumption than fully tuned digital architectures \cite{Chen2016} without compromising on generalization capabilities \cite{Huang2012}.The power consumption comes to $\approx$ $0.71$ $\mu$W. Thus, combining both decoder and feature extraction blocks merely adds up to $4.71$ $\mu$W to the total power consumption. This is three orders of magnitude lesser than the pre-conditioner circuits typically consuming power in the mW range \cite{Gao2012}. To summarize, we have shown that incorporating feature extraction and neuromorphic decoder blocks in the implant leads to up to three orders of magnitude reduction in the power intensive data rate, while adding an almost negligible fraction to the existing power budget. Accordingly, we hope for a significant increase in the battery lifetime \cite{Dethier2013}.

\begin{figure*}[t]
\centering
\includegraphics[scale=0.45]{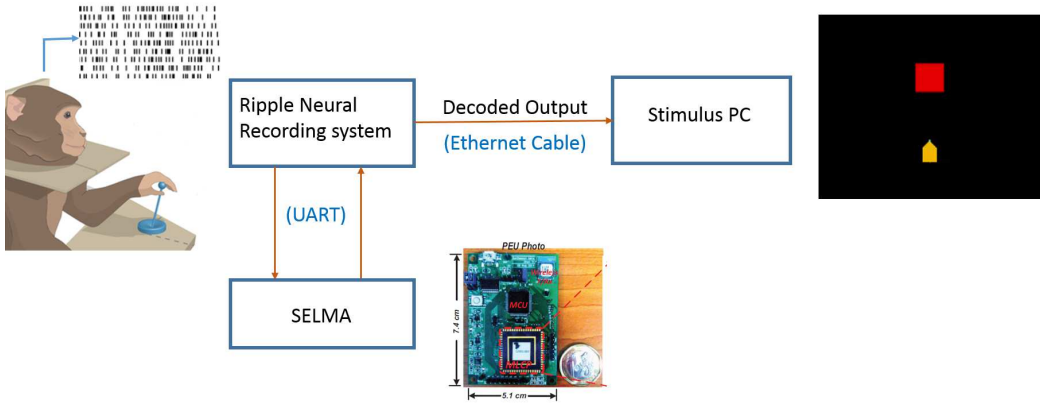}
\caption{Experimental setup for closed loop decoding with SELMA. SELMA and NHP figures are adapted from \cite{Chen2016} and \cite{Libedinsky2016b} respectively.}
      \label{fig:RealTimeSetup}
\end{figure*}

Before integrating blocks up to the proposed neuromorphic decoder in the implant itself, it is imperative to test if these blocks yield an acceptable performance in a closed loop real-time setting. This is the object of this work and we present a test set up in Fig. \ref{fig:RealTimeSetup} wherein the feature extraction and neuromorphic decoder blocks are currently being implemented externally. 

\section{SELMA - A NEUROMORPHIC DECODER}

\subsection{Algorithm}

SELMA \cite{Chen2016} is a hardware implementation of a randomized neural network such as the extreme learning machine (ELM) algorithm \cite{Huang2012, Huang2006}. To put it briefly, ELM is a single hidden layer feedforward neural network, where the first layer weights are random and fixed, and only the second layer weights are learned.

\begin{figure}[h]
\centering
\includegraphics[scale=0.7]{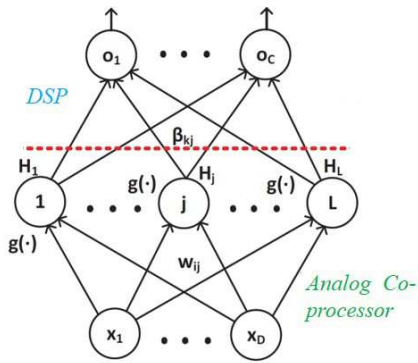}
\caption{Architecture of ELM neural network.}
      \label{figurelabel}
\end{figure}

\begin{flushleft}
The hidden layer is expressed as,
\end{flushleft}
\begin{IEEEeqnarray}{CCl}
$$H_j = g(\sum_{i=1}^{D}w_{ij} X_i + b_j)$$
%a & = & b + c
\nonumber\\
b_j,w_{ij} \ \epsilon \ \mathbb{R}; \ X_i \ \epsilon \ \mathbb{R}^D
\end{IEEEeqnarray} where $H_j$ is output of $j^{th}$ hidden layer neuron, $g(.)$ is the activation function, $w_{ij}$ are the input layer weights, $X_i$ is the input feature vector and $b_j$ is the bias for the respective hidden node. Input feature vector $X_i$, in case of spike based motor decoding studies, comprises of instantaneous firing rates $r_k(i)$ appearing at $k$ input electrodes at time $t=i$. 

\begin{IEEEeqnarray}{CCl}
X_i = \begin{bmatrix} r_1(i) & r_2(i) & \cdots & r_{D-1}(i) & r_D(i) \end{bmatrix}   
%a & = & b + c
\end{IEEEeqnarray}

\begin{flushleft}
Firing rates $r_k(i)$ are computed as the number of spikes appearing in a look back time window $T_w$ and are updated every $T_s$ seconds, such that $T_s = 1/(d_f)$, where $d_f$ is the decoder frequency.
\end{flushleft}

\begin{flushleft}
Final output $o_k$ at the $k^{th}$ output neuron is computed as,
\end{flushleft}

\begin{IEEEeqnarray}{CCl}
o_k = \sum_{j=1}^{L}\beta_{kj}H_j
%a & = & b + c
\end{IEEEeqnarray}where $\beta_{kj}$ represents the second layer weights.

\begin{flushleft}
Computation of the output weight involves solution of the following equation:
\end{flushleft}
\begin{IEEEeqnarray}{CCl}
\beta = H^{\dagger}T   \nonumber\\
T \ \epsilon \ \mathbb{R}^C;\ H\ \epsilon\ \mathbb{R}^L
\end{IEEEeqnarray}
where $T$ is the target vector and $C$ is the number of classes for a classification problem.

\subsection{Hardware blocks}

SELMA consists of two main blocks - a) An analog co-processor and b) DSP block. The analog co-processor has been custom designed employing sub-threshold analog design techniques for ultra-low power consumption \cite{Chen2016} whereas the DSP block is implemented on a low power micro-controller unit or MCU (TI-MSP430).

\begin{figure}[h]
\centering
\includegraphics[scale=0.45]{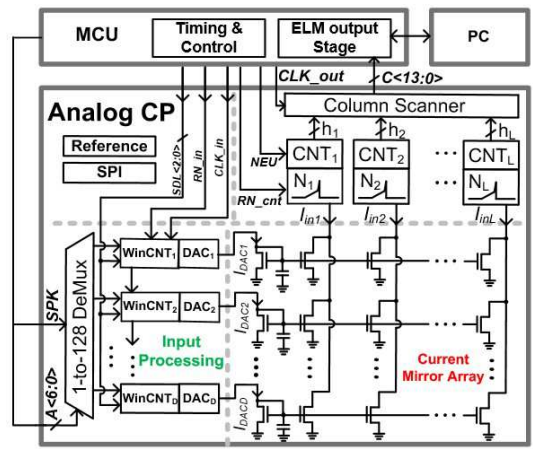}
\caption{Chip architecture for SELMA (modified from \cite{Chen2016}).}
      \label{figurelabel}
\end{figure}

The decoder takes spikes as inputs as they occur at the respective electrodes. The input processing circuit then computes the average firing rate for every channel based on a user defined time bin of $T_w$ seconds and updates it every $T_s$ seconds. This is followed by the current mirror array which exploits the inherent mismatch of transistors to implement the first layer of ELM in a power and area-efficient manner. The hidden nodes are implemented in the form of current controlled oscillators, which integrate the summed currents and yield an output every $T_s$ seconds. The hidden layer outputs are then multiplied by the weights stored in TI-MSP430 to yield the decoded outputs every $T_s$ seconds.

\section{METHODS}

\subsection{Neural Signal Acquisition and Signal Processing}
Two floating microwire arrays (32 electrodes each) were implanted in an NHP (Macaca fascicularis) in the left primary motor cortex area. Ripple$'$s grapevine neural recording system \cite{Ripple} was used for real-time recording of raw neural data sampled at $30$ kHz. Spikes were detected using median absolute deviation method \cite{Quiroga2004} with the threshold set as negative of five times the median.

\subsection{Behavioural Task}
An NHP is trained to maneuver a joystick using his left hand to drive a virtual wheelchair avatar towards a square target appearing on a computer screen in the hand-controlled joystick mode. In a given trial, wheelchair avatar starts at the centre of the screen and the target is presented in a pseudo-random manner in one of the three different directions $-$ Forward, right and left in the form of a square of side 2 cm. The wheelchair avatar is driven in a discrete control fashion at a frequency of $10$ Hz with possible position updates being one of left, right, forward or stop depending on the position of the joystick. A trial is considered successful if the NHP reaches, and stays in the target area for $1.5$ seconds under a total elapsed time of $13$ seconds. Once the NHP was trained well enough to perform the trials consistently, we proceeded to the neural control mode wherein the joystick is removed and the avatar is driven by the decoded outputs at 10 Hz. It has been observed that in the absence of joystick, the NHP makes little or no overt movements. 

\subsection{Decoder Training}

The decoder was trained on a total of three sessions consisting of $20$ trials each. The NHP was made to passively observe a series of successful trials for the first couple of sessions following the training paradigm in {\cite{Gilja2012}}. Thereafter, an intermediate model $M_{int}$ was trained based on these sessions. In the third session, $M_{int}$ was employed with $50$\% assistance \cite{Libedinsky2016b, Sadtler2014} in the system and the final trained model $M_f$ was obtained including data from all three sessions. Only successful trials were used for training the model.

\section{RESULTS}

The trained model $M_f$ was tested until the NHP performed $60$ successful trials each day by neural control. This experiment requires at least $\approx$ $70$\% \cite{Libedinsky2016b} of the decoded directions to match ground truth for a trial to be successful. 

\begin{figure}[h]
\centering
\includegraphics[scale=0.25]{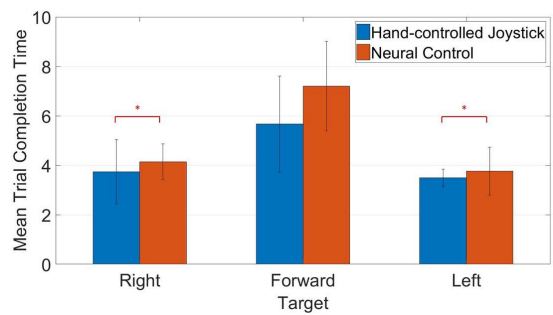}
\caption{Mean times to complete trials are comparable for closed-loop decoding by SELMA and hand controlled joystick experiments. Statistically insignificant difference is denoted by (*), p$>$0.01 (unpaired two-sample t-test)}
      \label{fig:result1}
\end{figure}

Chance level accuracy for the trained classifier model is $25$\%, so the chance level of success of a trial can be computed as $0.25^{(0.7\times130)}\times100 \approx  0$\%. Closed-loop decoding results are benchmarked against those obtained from hand-controlled joystick experiments. We have compared a total of $180$ hand-controlled joystick and neural decoder controlled closed loop trials over a period of $3$ days with $60$ trials performed on each day. 

Mean successful trial completion speeds for forward, left and right targets by brain control have been shown to be $\approx$ 79\%, 93\% and 90\% respectively of those obtained by hand-controlled joystick. The results are plotted in Fig. \ref{fig:result1}.
    
\begin{figure}[h]
\centering
\includegraphics[scale=0.25]{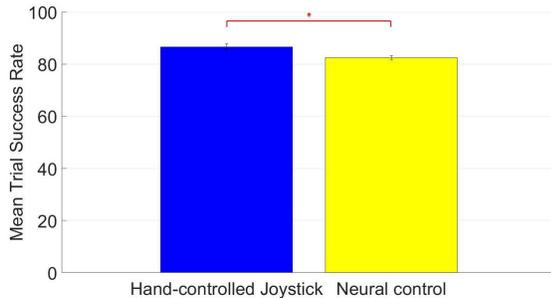}
\caption{Percentage of successful trials for hand-controlled joystick and neural control. Statistically insignificant difference is denoted by (*), p$>$0.01 (unpaired two-sample t-test)}
      \label{fig:result2}
\end{figure}

Mean success rates for closed-loop brain control is $\approx 96\%$ of that obtained by hand-joystick control as shown in Fig. \ref{fig:result2}.

\section{CONCLUSION}

From infection mitigation and mobility standpoints, a fully implantable iBMI is the one plausible approach. Neuromorphically engineered iBMI systems have long been touted as a viable option, albeit without hardware-based demonstration in standard NHP based experimental setups. We have demonstrated for the first time fast and accurate neural control obtained via a neuromorphic chip decoder in a closed loop NHP based experimental setup. The decoding accuracies are similar to those obtained in hand-controlled joystick experiments paving the way for future implanted system development.


%%%%%%%%%%%%%%%%%%%%%%%%%%%%%%%%%%%%%%%%%%%%%%%%%%%%%%%%%%%%%%%%%%%%%%%%%%%%%%%%

%%%%%%%%%%%%%%%%%%%%%%%%%%%%%%%%%%%%%%%%%%%%%%%%%%%%%%%%%%%%%%%%%%%%%%%%%%%%%%%%

%%%%%%%%%%%%%%%%%%%%%%%%%%%%%%%%%%%%%%%%%%%%%%%%%%%%%%%%%%%%%%%%%%%%%%%%%%%%%%%%
%\section*{APPENDIX}


\section*{ACKNOWLEDGMENT}
The authors would like to thank Abdur Rauf for helping in animal training and data collection.

%%%%%%%%%%%%%%%%%%%%%%%%%%%%%%%%%%%%%%%%%%%%%%%%%%%%%%%%%%%%%%%%%%%%%%%%%%%%%%%%

\bibliographystyle{IEEEtran}
%\bibliography{library.bib}

\end{document}